# Tunable microwave absorption performance of dual doped graphene by varying doping sequence

L. Quan[§], H.T. Lu[§], F.X. Qin[*,1], D. Estevez, Y.F. Wang, Y.H. Li, Y. Tian, H. Wang, H.X. Peng[**,2]

*Institute for Composites Science Innovation (InCSI)*

*School of Materials Science and Engineering*

*Zhejiang University*

*Zheda Road 38*

*Hangzhou, 310027, PR. China*

[§]: L. Quan and H.T. Lu contributed equally to this manuscript.

[*]: Corr author: Tel: 187 6718 9652. E-mail: faxiangqin@zju.edu.cn (F.X. Qin)

[**]: Corr author: Tel: 137 3800 6020. E-mail: hxpengwork@zju.edu.cn (H.X. Peng)

**Abstract**

Sulfur and nitrogen dual doped graphene have been extensively investigated in the field of oxygen reduction reaction, supercapacitors and batteries, but their magnetic and absorption performance have not been explored. Besides, the effects of doping sequence of sulfur and nitrogen atoms on the morphology, structural property and the corresponding microwave absorption performance of the dual doped graphene remain unexplored. In this work, nitrogen and sulfur dual doped graphene with different doping sequence were successfully prepared using a controllable two steps facile thermal treatment method. The first doping process played a decisive role on the morphology, crystal size, interlayer distance, doping degree and ultimately magnetic and microwave absorption properties of the dual doped graphene samples. Meanwhile, the second doping step affected the doping sites and further had a repairing or damaging effect on the final doped graphene. The dual doped graphene samples exhibited two pronounced absorption peaks which intensity was decided by the order of the doping elements. This nitrogen and sulfur dual doped graphene with controlled doping order provides a strategy for understanding of the interaction between nitrogen and sulfur as dual dopants in graphene and further acquiring microwave absorbing materials with tunable absorption bands by varying the doping sequence.

## 1. Introduction

The development of high-performance microwave absorption materials has become a recent focus due to the rapid increasing of electronic communication and the associated electromagnetic radiation which is harmful to biological systems.[1-3] Ideal microwave absorbing materials possess low thickness, low density and strong absorption over a broad frequency range are extremely favored.[4]

Owing to the advantages of low density and excellent electrical and thermal conductivities, graphene and graphene based materials have been extensively explored for microwave absorption applications.[5] Through the incorporation with dielectric and magnetic materials[6-8] or doping with nitrogen,[9-11] the absorption performance of graphene can be significantly improved. The latter approach has the advantage of not compromising the light nature of graphene while improving its magnetic property and thus ameliorating the impedance match characteristic. Sulfur doped graphene has been investigated in electrocatalyst,[12, 13] sensors,[14] low temperature magnetism,[15-17] and electromagnetic shielding.[18] However, it is worth noting that the sulfur doped graphene has been scarcely explored as a potential microwave absorber.[19] As for the dual doped graphene samples such as nitrogen&boron or nitrogen&sulfur co-doped graphene, they have been used for catalyst,[20] supercapacitors,[21] Li storage,[22] glucose biosensor[23] and electrocatalysts for oxygen reduction reactions.[24-26] However, it's worth noting that their application on microwave absorption remains unexplored, and no research has been done on tuning the properties of the co-doped graphene by varying the doping sequence of doping elements.

Most recently, the concept of "plainification" has been applied in both carbon and metallic materials, which indicates tunable properties can be achieved by precise structure design without compositing with other materials.[27-29] In this work we adopt the two steps heat treatment method to synthesize the dual doped graphenes with different doping sequence of nitrogen and sulfur elements. A schematic diagram of the synthesis of dual doped graphenes is arranged in **Figure 1**. The effect of different elements doping sequence on the morphological, structural, magnetic and microwave absorbing properties of the dual doped graphenes were investigated.

## 2. Experimental Details

### 2.1 Materials

Graphene oxide (GO) powder was purchased from C6G6 Technology Co., Ltd, China. Benzyl disulfide, urea and paraffin were purchased from Sinopharm Chemical Reagent Co.,Ltd. All the reagents and materials were used without further purification.

### 2.2 Synthesis of nitrogen and sulfur dual doped graphene (NSrGO)

Nitrogen and sulfur dual doped graphene (NSrGO) was prepared by grinding nitrogen doped graphene-NrGO (detailed NrGO preparation is put in Supplementary Information) with benzyl disulfide (BD) at a mass ratio of 1:10 and further heat treated at 270°C and 600°C each for 1 h. NSrGO represents the GO doped first with nitrogen and subsequently with sulfur. The schematic diagram is illustrated in Figure 1.

### 2.3 Synthesis of sulfur and nitrogen dual doped graphene (SNrGO)

Sulfur and nitrogen dual doped graphene (SNrGO) was prepared by grinding sulfur doped graphene SrGO (detailed preparation is explained in Supplementary Information) with urea at a mass ratio of 1:10 and further annealing at 170°C and 600°C each for 1 h. SNrGO stands for the graphene oxide doped first with sulfur and afterwards with nitrogen.

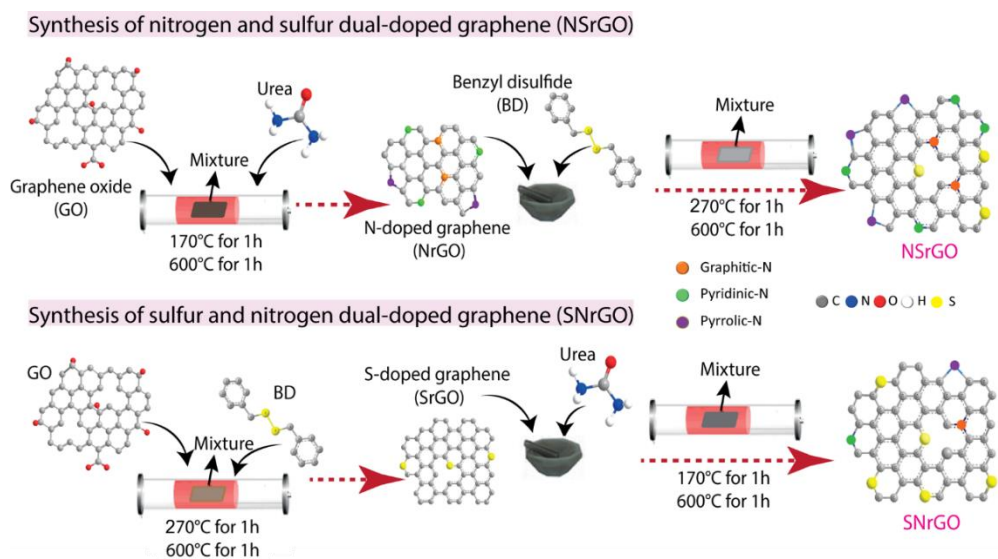

Figure 1 The schematic diagram of the synthesis for NSrGO and SNrGO.

## 2.4 Characterization

S-4800 cold field emission scanning electron microscope (FE-SEM) provided by Hitachi (Japan) was used to observe the morphology of the samples. Doping degree and chemical states of the doped elements were studied by AXIS SUPRA X-ray photoelectron spectrometer (XPS) supplied by Kratos Analytical Ltd.. DXR Smart Raman spectrometer (illumination wavelength of 532 nm) was used to examine the Raman spectrum. The XRD test was conducted on Rigaku SmartLab High Power X-Ray Diffractometer. Magnetic properties of the samples were tested using a Quantum Design Physical Measurement Property System - Vibrating Sample Magnetometer (PPMS-VSM). Scattering parameters (S-parameters) were measured through coaxial line method in the frequency range of 2~18 GHz using the R&S ZNB20 vector network analyzer (VNA). The sample used for coaxial line test was prepared by mixing paraffin with the sample at a mass ratio of 8:2, then the mixture was pressed into a standard toroidal shape (inner diameter: 3.04 mm, outer diameter: 7 mm). According to the measured S-parameters (S11, S21, S12, S22), the complex permittivity ($\varepsilon_r$) and permeability ($\mu_r$) can be calculated through the NRW (Nicolson-Ross-Weir) method,[31] and the reflection loss (RL) can be calculated with electromagnetic parameters through transmission line theory [32]

$$Z_{in} = Z_0\sqrt{\mu_r/\varepsilon_r}\tanh\left[j\left(2\pi fd/c\sqrt{\varepsilon_r/\mu_r}\right)\right] \quad (1)$$

$$RL = 20\log\left|\frac{(Z_{in}-Z_0)}{(Z_{in}+Z_0)}\right| \quad (2)$$

where $c$ is the speed of light in vacuum, $f$ is the frequency of the incident electromagnetic wave, $d$ is the thickness of the absorbing material, $Z_0$ is the impedance of the free space, and $Z_{in}$ is the input impedance of the absorbing material. Therefore, the closer the values of $Z_{in}$ and $Z_0$ means the better impedance match, the less wave will be reflected. s

## 3. Results and Discussions

Initially, the influence of different annealing temperatures and varied GO/BD weight ratios on the morphology, structure and absorption performance of sulfur single doped graphene (SrGO) was investigated (Supplementary Information). According to the macro powder morphology of the SrNO and NrGO, at the same weight, the volume occupied by the NrGO sample was significantly larger than that of the SrGO sample, which means the bulk density of the NrGO was smaller than SrGO. The micromorphology difference of GO, rGO, SrGO, NrGO as well as the dual doped graphene samples NSrGO and SNrGO are illustrated in **Figure 2**. GO as the starting precursor for all the followed samples shows a crumpled appearance; rGO exhibits an stacked structure after the releasing of most of the oxygen containing groups; SrGO still looks the same with GO, and Fig.2c-f illustrate heating temperatures independent phenomena of SrGO appearance; NrGO shows a much different morphology from GO. As for the morphologies of dual doped graphene NSrGO and SNrGO, if we comparing the morphology with their mid-products, we can find that they both share similarities to their mid-product NrGO and SrGO. Pronounced restacking phenomena are observed in the nitrogen first doped samples (NrGO, NSrGO), but sulfur first doped samples (SrGO and SNrGO) remain unchanged to their precursor GO. Such morphological differences implying that the doping level and reduction degree of NrGO and NSrGO might be much higher than that of SrGO and SNrGO. These observations indicate that the morphology of the dual doped graphene is determined by the first doping step and hardly being affected by the second doping process.

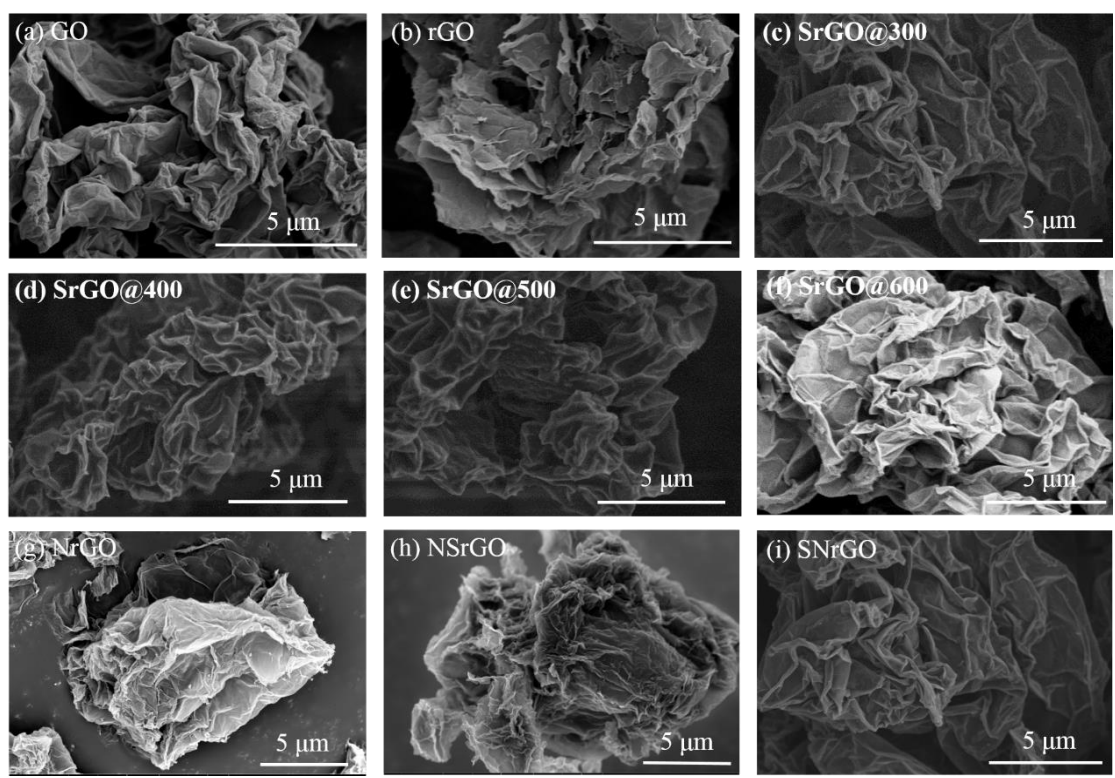

**Figure 2** Morphological features of GO (a); rGO (b); SrGO at different preparation temperatures, 300°C (c), 400°C (d), 500°C (e), 600°C (f); NrGO (g), NSrGO (h) and SNrGO (i).

According to the Raman results in **Figure 3**a, the D, G and 2D peaks are observed respectively, which are characteristic peaks of graphene materials. From the point of view of peak intensities, the SrGO and SNrGO exhibit much higher intensity and relatively narrow peaks than the NrGO and NSrGO samples. In addition, the 2D peak disappears from NrGO and NSrGO spectrum, which is a result of excessive doping in the nitrogen first doped samples.[33] The slightly blue-shift of the G peak also confirms the high doping degree of NrGO sample compared to the rest of the samples.[33] In contrast, the better preserved graphitic structure and less doping degree of SrGO and SNrGO samples account for the presence of 2D peak in their Raman spectra. Regarding the $I_D/I_G$ ratio which is an indication of defects,[34] the $I_D/I_G$ ratio of SrGO is much smaller than that of NrGO (also lower than that of GO and rGO, Fig.S6); while it slightly increases in SNrGO sample with further doping SrGO with nitrogen, indicating a damaging effect of nitrogen doping. On the other hand, further doping of NrGO with sulfur results in a significantly reduced $I_D/I_G$ ratio in NSrGO sample. Therefore, we can conclude that nitrogen doping results in an increase of $I_D/I_G$ ratio, thus a destructive effect to the graphene structure; doping with sulfur leads to a decrease in $I_D/I_G$ ratio, thence a certain recovery effect to the graphene structure. Not only for the single doped graphene but also in the dual doped graphene samples. The crystal size $La$ can be calculated from $La = \lambda * IG/ID$,[35] and it follows the order SrGO > SNrGO > NSrGO > NrGO. These results further prove that different doping sequences of the doping atoms can pose different effects on the intrinsic structure of graphene.

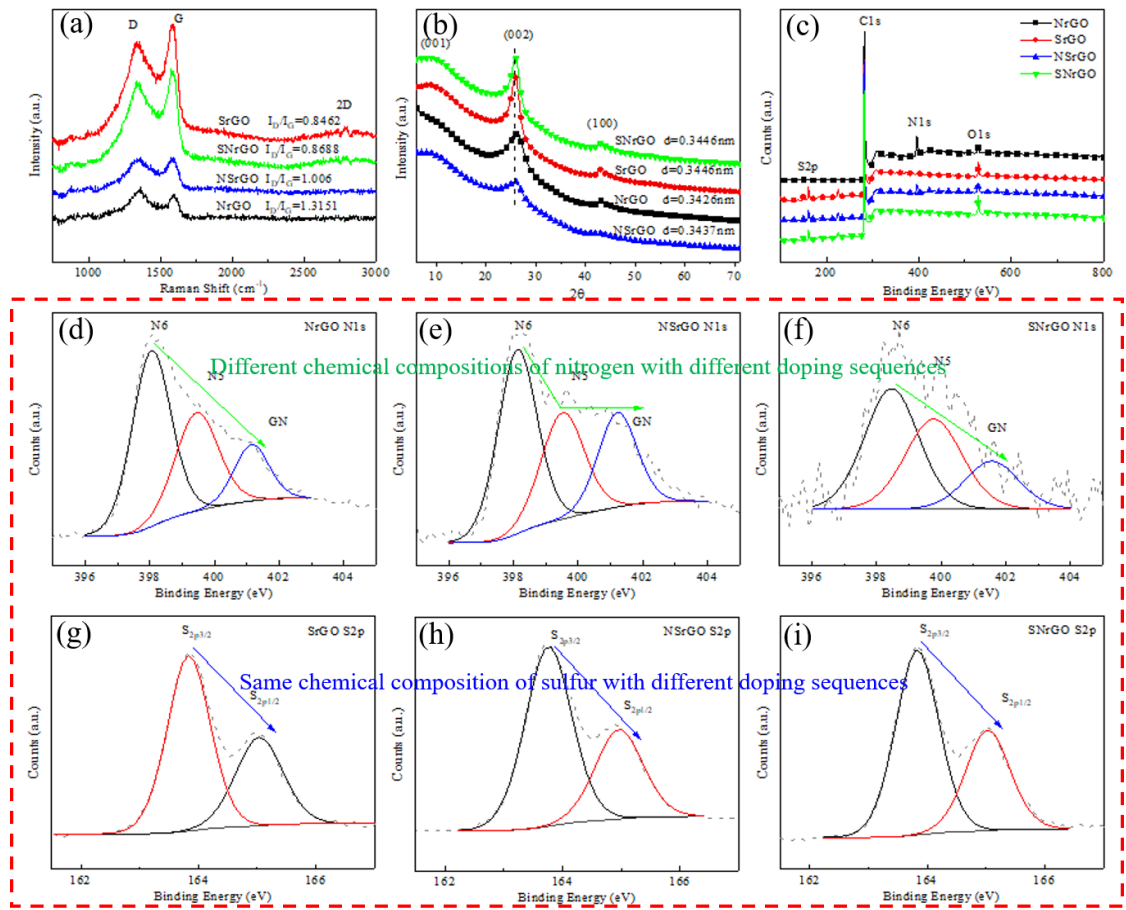

**Figure 3** The Raman (a), XRD (b), XPS (c) spectrum; and high resolution N1s spectra of NrGO (d), N1s and S2p for NSrGO (e,f), S2p of SrGO (h), (g,i) N1s and S2p of SNrGO sample.

The XRD patterns of the samples in the range of 7~70° are shown in Fig.3b. The characteristic (001) peak completely disappears in NrGO sample but a broad (002) peak exhibits at 2θ=26.1° corresponding to a distance of 0.3426 nm ($d = n\lambda/2 \sin\theta$) between NrGO layers [36]. As for SrGO, a small (001) peak is visible around 10.8°, and a relative narrow peak at 25.9° appears, implying a larger interlayer distance with respect to NrGO. In terms of the second doping step, i.e. doping NrGO with sulfur (NSrGO), an increased distance between the layers was found which is attributed to the large atomic diameter of sulfur. In contrast, further doping SrGO with nitrogen (SNrGO) has negligible influence on the layer distance. As a consequence, by varying the doping order of the nitrogen and sulfur, NSrGO and SNrGO will possess different interlayer distances. These results demonstrate the impact of doping sequence on the structural property of the dual doped graphene.

To analyze the influence of doping sequence on the elemental composition and chemical status, the XPS analyzation and high resolution XPS of N1s and S2p are shown in Fig.3c-i. The C1s, O1s, N1s peaks and C1s, O1s, S2p peaks can be found in NrGO and SrGO spectra, respectively; while both N1s and S2p are present in the NSrGO and SNrGO spectra (Fig.3c), implying the successful doping of nitrogen and sulfur in the dual doped samples. In addition, the intensity of O1s peak in NrGO and NSrGO is weaker than SrGO and SNrGO, indicates a lower oxygen containing groups content of the nitrogen first doped samples. For single doped graphene, the N/C percentage calculated is 9.9 at.% for NrGO and 3.8 at.% S/C for SrGO (elemental compositions are arranged in Table 1). Three kinds of nitrogen configurations are elucidated from the N1s spectra of NrGO (Fig.3d): pyridinic-N at 397.8 eV, pyrrolic-N at 399.8 eV and graphitic-N at 401.3 eV.[36] From the S2p XPS spectra of SrGO (Fig.3g), two kinds of sulfur configurations S2p 3/2 at 163.7 eV and S2p 1/2 at 164.9 eV are recognized. C-SOx is not found in our sample owing to the high heat treated temperature of 600°C which avoids this functional group.[18, 37] Further doping NrGO with sulfur, i.e. NSrGO, the N/C percentage is decreased to 4.7 at.% and 2.2 at.% of S/C is obtained. After sulfur doping, the percentage of three kinds of nitrogen sites in NSrGO (Fig.3e) are different from the initial NrGO (Table S1). The percentages of pyridinic-N and pyrrolic-N are decreased while graphitic-N in NSrGO has a slightly increase in comparison to NrGO (Fig.3e and Table 1), indicating that sulfur is mainly introduced through the pyridinic-N and pyrrolic-N sites. Interestingly, the sulfur composition of NSrGO provided with S2p 3/2 and S2p 1/2 is quite similar to SrGO (Fig.3h and Table 1). Regarding the nitrogen doped SrGO, SNrGO possesses 2.9 at.% S/C and only 1.51 at.% N/C implying that the sulfur doped graphene cannot be further heavily doped due to the low doping degree of SrGO. For SNrGO sample, from the perspective of nitrogen and sulfur elemental composition, the relative content of sulfur sites S2p 3/2 and S2p 1/2 shown in Fig.3i are very stable, while the three nitrogen components (Table 1) are slightly different from the NrGO and NSrGO. These XPS elemental composition results show that sulfur can still be incorporated into highly nitrogen-doped graphene (NrGO), while nitrogen can be barely introduced into the sulfur doped graphene (SrGO). In summary, diverse doping order of nitrogen and sulfur result in different nitrogen distribution sites and doping degree of each element. These results demonstrate that the doping sequence has a significant influence on the doping degree and the composition of different elements.

Table 1 The elemental composition of NrGO, SrGO, NSrGO and SNrGO.

|  | N/C (at. %) | S/C (at. %) | Pyridinic N ratio (%) | Pyrrolic N ratio (%) | Graphitic N ratio (%) | S2p 3/2 ratio (%) | S2p 1/2 ratio (%) | Pyrrolic N (%) = N/C * Pyrrolic N ratio |
|---|---|---|---|---|---|---|---|---|
| NrGO | 9.9 |  | 52.9 | 32.1 | 15.0 |  |  | 3.18 |
| SrGO |  | 3.8 |  |  |  | 65 | 35 |  |
| NSrGO | 4.7 | 2.2 | 46.3 | 29.7 | 24.0 | 66.5 | 33.5 | 1.39 |
| SNrGO | 1.51 | 2.9 | 46.1 | 35.8 | 18.1 | 64.3 | 35.7 | 0.54 |

The magnetic property of nitrogen or sulfur single doped graphene has been widely studied in the literature[16, 38-40], while the effect of doping sequence on magnetic property of dual doped graphene remains unexplored. The hysteresis loops of nitrogen and sulfur single doped graphene, as well as dual doped graphene with different doping order are shown in **Figure 4**. All the samples achieved the saturation magnetization (Ms) at 60000 Oe, and their Ms follows the order NrGO > NSrGO > SNrGO > SrGO. This order agrees with that of $I_D/I_G$ ratio, while it is opposite to the crystal size *La* order. To explain such a trend of Ms, three aspects must be considered. The first one is related with the doping degree, according to the first principle calculation it shows that each doping atom can introduce certain magnetic moments,[41] and in general the higher of the doping degree the larger of the Ms. This is consistent with the $I_D/I_G$ ratio and the XPS results. The second aspect is related to defects like edges, as Shenoy et al. demonstrated, edge states (zigzag, armchair edges) of graphene can trigger different magnetic behavior.[42] Recalling the Raman results, the different $I_D/I_G$ ratios of samples result in different crystal sizes. For the same content of sample, a larger crystal size means a lower percentage of edges, which results in a lower Ms. The third one is their content of pyrrolic N in those samples (calculated as N/C*Pyrrolic N from Table 1) which plays the most important role in the contribution to magnetic moments.[11] Except for their Ms difference, the magnetic responding rates to the external field (magnetic susceptibility, expressed as the slope at the origin in Fig.4, k1 and k2) of the nitrogen first-doped graphene samples (NrGO, NSrGO) and sulfur first doped graphene samples (SrGO, SNrGO) are different; the magnetic susceptibility of the nitrogen-first doped graphenes are larger than the sulfur-first doped graphene samples due to the considerable larger doping degree and higher pyrrolic N concentration. The magnetic performances of the four samples verified the substantial influence of different doping elements and their doping order.

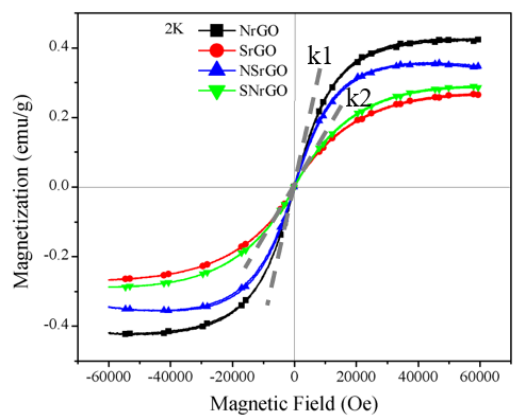

**Figure 4** The hysteresis loop of NrGO, SrGO, NSrGO and SNrGO (slope of k1 and

k2 represent the magnetic response rates of the samples to external field) at 2K.

The dielectric spectrum can provide important information about the polarization mechanism which is closely related with the microwave attenuation and absorption, thus it is an important characteristic of a microwave absorbing material. To obtain the permittivity of a material, generally, the sample is dispersed in wave-transparent material to obtain a composite; then the S-parameters of the composite can be analyzed with VNA and hereafter permittivity can be calculated from S-parameters with the NRW method. **Figure 5**a-b are the SEM images of 5 wt.% NSrGO and SNrGO dispersed in polydimethylsiloxane (PDMS). It can be seen from the figures that at the same loading content, the dispersibility of NSrGO is significantly better than that of SNrGO, which is closely related with their bulk density. As we mentioned above, NrGO holds a lower bulk density than SrGO, and owing to the first doping decisive effect, the bulk density of NSrGO is also smaller than SNrGO, which makes NSrGO own a better dispersibility than NSrGO. The dispersibility of the sample in the matrix will further affect dielectric spectrum of the composite. The permittivity of the dual doped graphene samples with different doping sequence are illustrated in Fig.5c-d. NSrGO possesses a much higher complex permittivity and a stronger frequency dependent feature than the SNrGO owing to its good dispersibility. Even though both nitrogen and sulfur doping are *n*-type doping and they are believed to enhance the conductivity of graphene through electron donation and increasing the carrier concentration,[43] the permittivity of SNrGO is merely slightly higher than GO.[9] Figure 5c-d are the cole-cole plots of the dual doped samples, both of them contain four different polarization relaxation processes. A schematic diagram on the polarization mechanism of the doped samples is shown in Fig.5g. The introduced nitrogen and sulfur atoms in the graphene lattice, which can act as tiny dipoles or polarized centers under the external field and further lead to polarization relaxation.[44] Such effect is more pronounced in NSrGO sample owning to its smaller crystal size and higher doping content compared to SNrGO sample, which leads to a higher permittivity of NSrGO than SNrGO. In addition, the dielectric mismatch between doped graphene and paraffin leads to entrapment of charges at the interfaces, resulting in interfacial polarization.[32, 43, 45] The good dispersibility of NSrGO make it possess more interfaces than SNrGO. Moreover, the trapped changes at the interfaces will cause the formation of large numbers of micro-capacitors. All the proposed polarization processes above will contribute to the permittivity difference.[46]

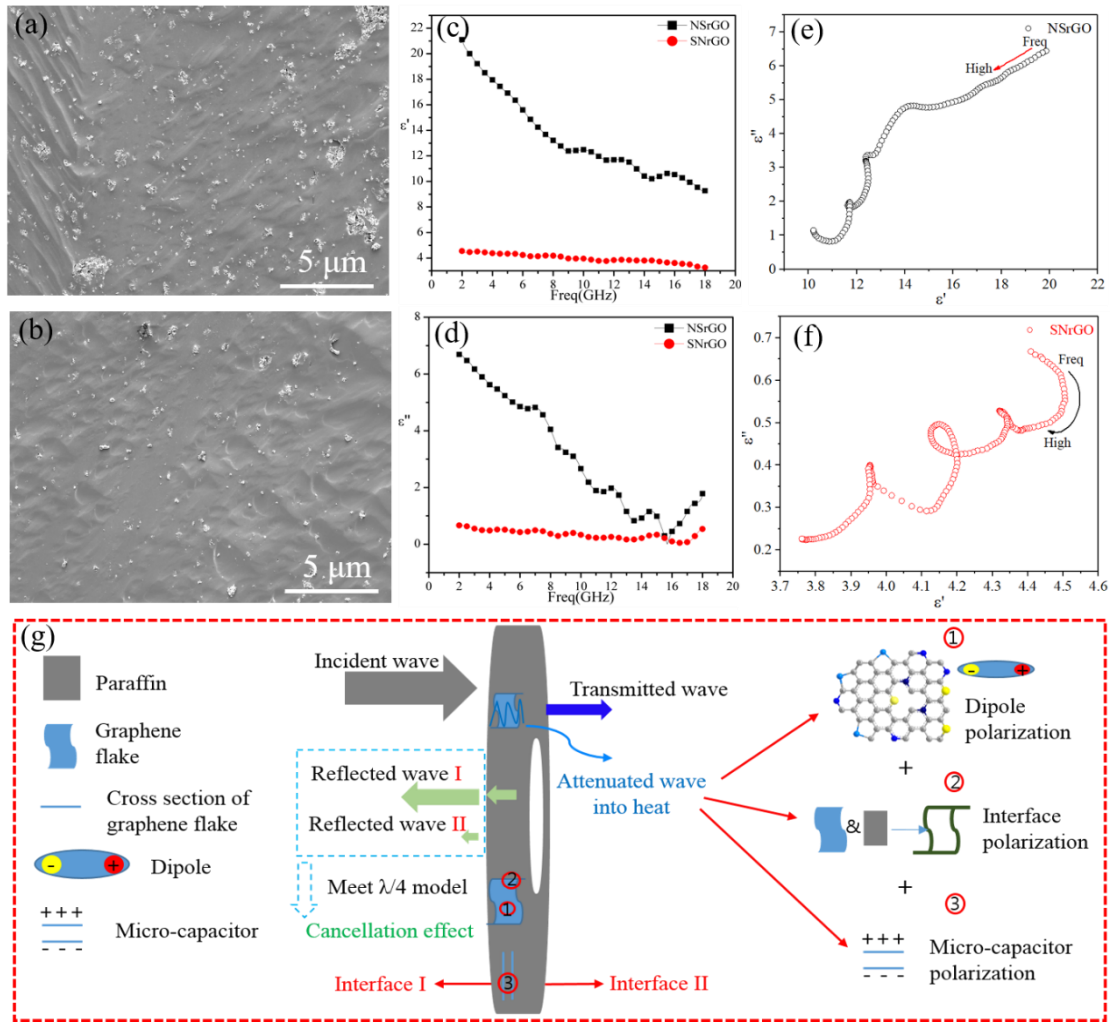

**Figure 5** SEM images of the dispersion of 5 wt.% NSrGO (a) and SNrGO (b) in PDMS matrix; the real (c) and image (d) part of permittivity; cole-cole plots of NSrGO (e) and SNrGO (f); schematic of the microwave absorption mechanism (g).

With respect to the microwave absorbing performance in **Figure 6**, according to our previous work, the absorption peaks of NrGO are at the range of 5~10 GHz (Fig.6a); and the absorption peaks of the sulfur doped graphene are at 11~16 GHz (SI, Section I and Section II). Regarding to the dual doped graphene samples (NSrGO and SNrGO samples), in Fig.6c-d, their RL curves tend to have two absorption peaks, the absorption peak at 5~10 GHz can be ascribed to the contribution of nitrogen doping; the absorption peak at 11~16 GHz of can be ascribed to the contribution of sulfur doping. The characteristic nitrogen peak is significantly deeper than sulfur characteristic peak in NSrGO. In contrast, the characteristic sulfur peak is clearly stronger than the nitrogen peak in SNrGO sample. In total, the absorption performance of SNrGO is superior at high frequency while NSrGO is valuable at lower frequency, which further demonstrates the dominant effect of the doping order. The total reflection loss values are not so outstanding as compared with our previous works, while the permittivity of the sample and thus absorption property can be easily tuned by adjusting the loading content of the sample in paraffin.[10]

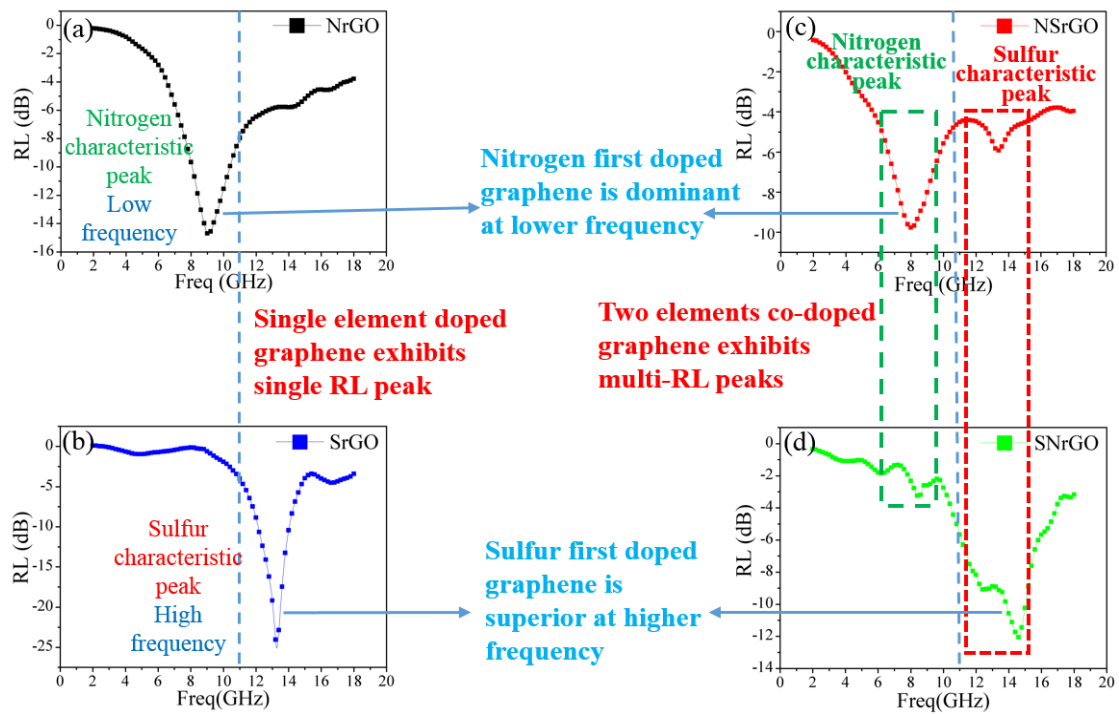

**Figure 6** The RL curves of NrGO (a), SrGO (b), NSrGO (c), SNrGO (d) at d=3.5 mm

## 4. Conclusions

In this paper, we have synthesized nitrogen and sulfur dual doped graphene through facile heat treatment method. By controlling the doping elements and doping sequence, dual doped graphene with different structures and performances were acquired. The first doping stage determined the morphology, doping degree, and restoration or deterioration effect to the dual doped graphene. The second doping step could further influence the doping sites of the resulted graphene, for instance, doping NrGO with sulfur resulted in more stable graphitic-N sites in NSrGO. The nitrogen and sulfur doping contents as a function of doping order is quite different, sulfur can be introduced through the less stable nitrogen sites while nitrogen is hard to be incorporated into the graphene sheets at the second doping step. Magnetic performance of the dual doped graphene also showed that the first doping stage is dominant on deciding the ultimate magnetic property of the dual doped graphene. Similar magnetic behavior was found between NSrGO and NrGO, as well as between SNrGO and SrGO. NSrGO and NrGO exhibited a higher Ms than SNrGO and SrGO owing to their higher doping degree and pyrrolic N content, respectively. Regarding the microwave absorption performance, the dual doped graphene displayed two typical absorption peaks (sulfur characteristic peak at 11~16 GHz and nitrogen characteristic peak at 5~10 GHz), sulfur first doped SNrGO was superior at high frequency while nitrogen first doped NSrGO was dominant at lower frequency. Overall, through modulating the doping elements and doping sequence of different elements, dual doped graphene samples with varied structure, magnetic and absorption properties were obtained, which shedding lights on the interaction between nitrogen and sulfur as co-dopants in graphene, as well as the acquisition of microwave absorbing material with tunable absorption bands by varying the doping sequence.

**Acknowledgements**

This work is supported by Aeronautical Science Foundation GFJG-112207-E11502, NSFC No. 51671171, 'National Youth Thousand Talent Program' of China and China Scholarship Council (CSC). We also appreciate use of the XRD instrument at the Institute for Basic Science Center for Multidimensional Carbon Materials.